# Graphene Transistor as a Probe for Streaming Potential

A. K. M. Newaz[†], D. A. Markov[¶,‡], D. Prasai[§], and K. I. Bolotin[†,*]

[†]Department of Physics and Astronomy, Vanderbilt University, Nashville, TN 37235, USA
[¶]Department of Cancer Biology, Vanderbilt University Medical Center, Nashville, TN 37232
[‡]Vanderbilt Institute for Integrative Bio-systems Research and Education (VIIBRE), Vanderbilt University, Nashville, TN 37235, USA
[§]Interdisciplinary Graduate Program in Materials Science, Vanderbilt University, Nashville, TN 37235, USA
* kirill.bolotin@vanderbilt.edu

ABSTRACT: We explore the dependence of electrical transport in a graphene field effect transistor (GraFET) on the flow of the liquid within the immediate vicinity of that transistor. We find large and reproducible shifts in the charge neutrality point of GraFETs that are dependent on the fluid velocity and the ionic concentration. We show that these shifts are consistent with the variation of the local electrochemical potential of the liquid next to graphene that are caused by the fluid flow (streaming potential). Furthermore, we utilize the sensitivity of electrical transport in GraFETs to the parameters of the fluid flow to demonstrate graphene-based mass flow and ionic concentration sensing. We successfully detect a flow as small as~70nL/min, and detect a change in the ionic concentration as small as ~40nM.

Recent advances in micro- and nano- fluidics have spawned a great interest in miniaturized nanoscale probes that can detect properties of liquid flowing through narrow channels. Field-effect transistors fabricated using nanotubes,[1, 2] nanowires,[3] and nanobelts[4] have been employed to sensitively detect mass flow, pH, and ionic strength of various fluids. Multiple chemical and biological applications of these devices ranging from acidity testing[5] to DNA sensing[6] have been demonstrated. However, while the majority of the experiments investigated the influence of stationary fluids onto electronic transport in nanoscale devices, there has been less progress towards developing nanoscale probes that can access the dynamics of the fluid flows. Diverse kinetic phenomena – such as Coulomb drag,[7] surface ion hopping,[8] phonon drag,[7, 8] fluctuating asymmetric potential,[2] and streaming potentials[1, 3, 9-11] – are associated with moving fluids, and can influence the conductance of a nanoscale device thereby making interpretation of the experimental results challenging.

Graphene, a single monolayer of graphite, is a novel nanoscale material that is uniquely suited for applications in fluidic sensing. Since every atom in graphene belongs to its surface, electron transport in graphene is expected to be exquisitely dependent on the disturbances caused by the liquid flow in the immediate micro- and nano-environment of graphene.

In addition, graphene holds several important advantages over other nanoscale materials such as carbon nanotubes or nanowires. First, the carrier density in graphene, unlike in nanotubes and nanowires, can be directly determined from the Hall voltage measured in an external magnetic field. This enables sensitive measurements of the electric fields in the immediate vicinity of graphene sheets. Second, the ability to fabricate graphene into any desired shape and size makes it more attractive compared to other nanoscale materials, such as nanotubes and nanowires, which are difficult to control at the nanoscale. Finally, the robust nature of carbon-carbon bonds in graphene makes it biocompatible and chemically stable, enabling multiple potential device applications.

Here, we investigate the influence of the fluid flow on electrical transport of graphene field effect transistors (GraFETs) placed inside a microfluidic channel. We find a large and reproducible shift of the charge neutrality point (CNP) of graphene that is proportional to the liquid velocity within the microfluidic channel. This shift depends on both the flow velocity and the concentration of ions in the liquid, and is interpreted as due to the streaming potential developed in an electrolyte flowing past dielectric surface of the channel. Furthermore, we employ the observed phenomena to design a graphene-based mass-flow and ionic strength sensors. The sensitivity of these label-free sensors is ∼300 times higher than the reported flow sensitivity of a device based on carbon nanotubes[1] and ∼4 times higher compared to a device based on Si nanowire.[9] In addition to the mass-flow sensing, we demonstrate that GraFETs can detect changes in the ionic strength of a moving liquid with the sensitivity ∼40nM.

In our experiments, we employed devices that contain one or two independently contacted graphene field effect transistors that are placed inside a single microfluidic channel (Fig. 1a). The fabrication started with either growing graphene via chemical vapor deposition (CVD) on copper foils and transferring it onto $SiO_2$(300nm)/Si substrate[12] or directly depositing graphene that is mechanically exfoliated from Kish graphite onto a similar substrate.[13] In both cases, the single layer character of graphene was confirmed using Raman spectroscopy.[14] Graphene was then patterned either into narrow (20μm×30μm) strips or into Hall-bar shaped devices with six probes. The devices were then contacted electrically using Cr/Au (2nm/80nm) electrodes deposited via thermal evaporation. The microfluidic channels (80 μm tall and 50 μm wide) were formed in polydimethylsiloxane (PDMS) using standard soft-lithography and replica molding techniques.[15,16] Once PDMS was cured and access holes were punched, it was placed onto the GraFET device on the Si substrate such that the transistor was in the middle of the microfluidic channel and clamped to form a leak-tight seal. Since the exposure to oxygen plasma is detrimental to the graphene,[17] the PDMS surface and the $SiO_2$ did not go through any oxygen plasma cleaning.

In a typical experiment, we examined the electrical transport properties of GraFETs as a function of the liquid ionic strength and flow velocity. Overall, we studied two multi-probe devices and four 2-probe devices. For Hall-bar shaped devices, we recorded longitudinal resistance ($R_{xx}$) at zero magnetic field ($B$=0T) and Hall resistance ($R_{xy}$) at $B$=±46mT as a function of the counterelectrode voltage $V_{CE}$ that was applied to a platinum needle placed 7mm downstream from the GraFET (Figs. 1a). This voltage was always kept within ±1V range to avoid hydrolysis of water and electrochemical modification of the electrodes. The flow of the liquid (a solution of NaCl in DI water) through a PDMS microfluidic channel was controlled by a syringe pump (Harvard Apparatus Pico Plus). The concentration of NaCl (ionic strength) was varied in the range of $N$=10 nM – 1 mM, and the volume flow rate $Q$ was between 0.07 and 300 μL/min. To allow for better comparison between the GraFET based devices studied here and previously reported experiments employing carbon nanotubes[1] and Si nanowires[3], we report the average flow velocity $u$ (rather than volumetric flow rate) that is calculated as $u=Q/A$, where $A$=4000 μm² is the cross-sectional area of the channel.

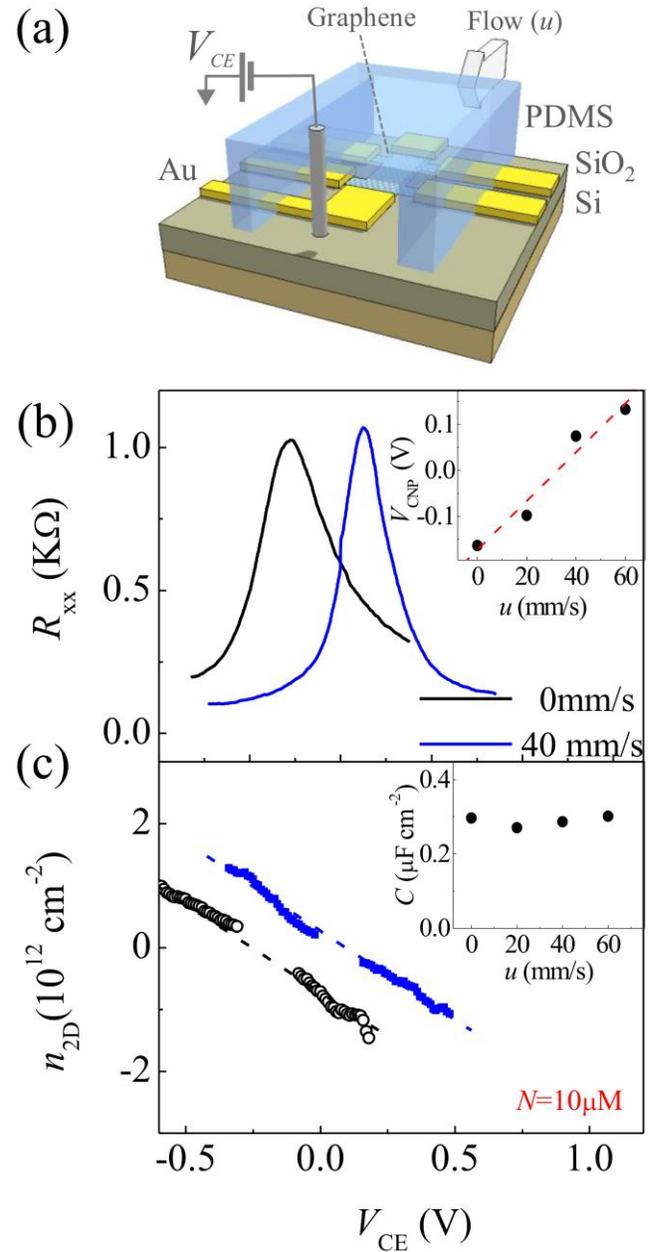

**Figure 1: The effect of the fluid flow on electronic transport in a 6-probe GraFET (Device #1). a)** Schematic representation of the experimental setup. **b)** The longitudinal resistance ($R_{xx}$) measured at $B$=0T as a function of the counterelectrode voltage $V_{CE}$. Black line corresponds to a stationary water solution of 10μM NaCl and blue line – to the flow of the same liquid at the average velocity $u$=40 mm/s. **Inset:** The position of the charge neutrality point $V_{CNP}$ for the same device vs. $u$. The dashed line is a fit to the data. **c)** The carrier density $n_{2D}$ extracted from the Hall resistance $R_{xy}$ vs. $V_{CE}$ for the same device. The carrier density data close to the CNP have been excluded due to the large non-uniformity in $n_{2D}$ on the either side of CNP[18]. **Inset**: The effective counterelectrode-graphene capacitance $C=edn_{2d}/dV_{CE}$ vs. average velocity $u$.

As a base-line experiment, we first examine the behavior of a GraFET placed in a stationary fluid and present the data from a typical Hall bar shaped device (device #1) fabricated using Kish graphite and placed in a stationary 10 μM solution of NaCl (Fig. 1a). We observed a sharp peak in the longitudinal resistance $R_{xx}$ as a function of the counterelectrode voltage $V_{CE}$ (Fig. 1b, black line), with the point of maximum resistance (CNP) located at the voltage $V_{CNP}$~-0.16 V. We found that the full width at half maximum (FWHM) of $R_{xx}(V_{CE})$ is quite small for every device evaluated, FWHM≤0.2V.

From the Hall resistance $R_{xy}(V_{CE})$ measured in a perpendicular magnetic field $B=\pm46$mT, we calculated the carrier density $n_{2D}=\Delta B/e\Delta R_{xy}$ (Fig. 1c, black curve) and the counterelectrode-graphene capacitance $C_g=edn_{2D}/dV_{CE}$~0.3 μFcm$^{-2}$ (Fig. 1c,Inset).

Both the large value of the gate capacitance and a very sharp peak in resistance have been previously reported for graphene devices measured in static ionic liquids and are the consequences of the so-called electrolyte gating.[19-22] When the voltage is applied between graphene and the counterelectrode, mobile ions present in the liquid are drawn towards the surface of the graphene forming an electric double layer (EDL).[11] As a result, the difference between local electric potential of the liquid $V_L$ (~$V_{CE}$ for the case of stationary liquid) and the potential of graphene (that is kept grounded) falls across the double layer of thickness $d$~$\varepsilon_0\varepsilon/C_g$~50-100 nm, where $\varepsilon_0$ is the vacuum permittivity and $\varepsilon$~80 is the static dielectric constant of water. The small thickness of the EDL and the large dielectric constant of water result in a large graphene-liquid capacitance, which is close to the graphene-counterelectrode capacitance.

When a fluid was set in motion by a syringe pump, we observed significant changes in the electrical transport in GraFETs: both the $R_{xx}(V_{CE})$ and $R_{xy}(V_{CE})$ curves shift with increasing flow velocity $u$ (Fig. 1b, blue curve corresponds to $u=40$ mm/s), with the value of the shift proportional to $u$ (Fig. 1b, Inset). Both the overall shape of these curves and the gate capacitance value (the slope of the $n(V_{CE})$ in Fig. 1c) are virtually unchanged (within experimental uncertainty) for the range of average fluid velocities employed in our experiments (Fig. 1c, Inset). We propose that the flow-speed-independent capacitance indicates that the structure of the EDL is not significantly affected by the flow of the liquid. At the same time, the flow-dependent shift of the CNP suggests that the local potential of the liquid near graphene depends on $u$ and is different from the potential of the counterelectrode.

The variation of the local potential of the liquid along the length of the channel is commonly observed in electrochemistry and referred to as 'streaming potential'.[23] Indeed, in contact with the electrolyte, walls of the channel (PDMS and SiO$_2$) acquire a net electrical charge.[3, 23, 24] To screen it, ions from the solution form an electrical double layer (EDL) next to the channel walls. When the flow through a microfluidic channel is initiated (pumps are turned on), the motion of the unbalanced charge in the diffuse part of the electrical double layer results in a net electrical current that is proportional to the flow velocity (Fig. 2c). The resulting redistribution of charge leads to the variation of the local potential along the direction of the flow. Therefore, the local potential of the liquid next to graphene is different from $V_{CE}$ by an amount $\Delta V$ that depends on the flow rate, counterelectrode-graphene distance, and concentration of ions in the liquid. We also argue that the contribution of other mechanisms, such as ion hopping[8], phonon drag,[7, 8] and fluctuating asymmetric potentials,[2] to the induction of electrical currents in graphene is negligible, since the liquid is nearly static near the surface of graphene for a pressure-driven microfluidic system.

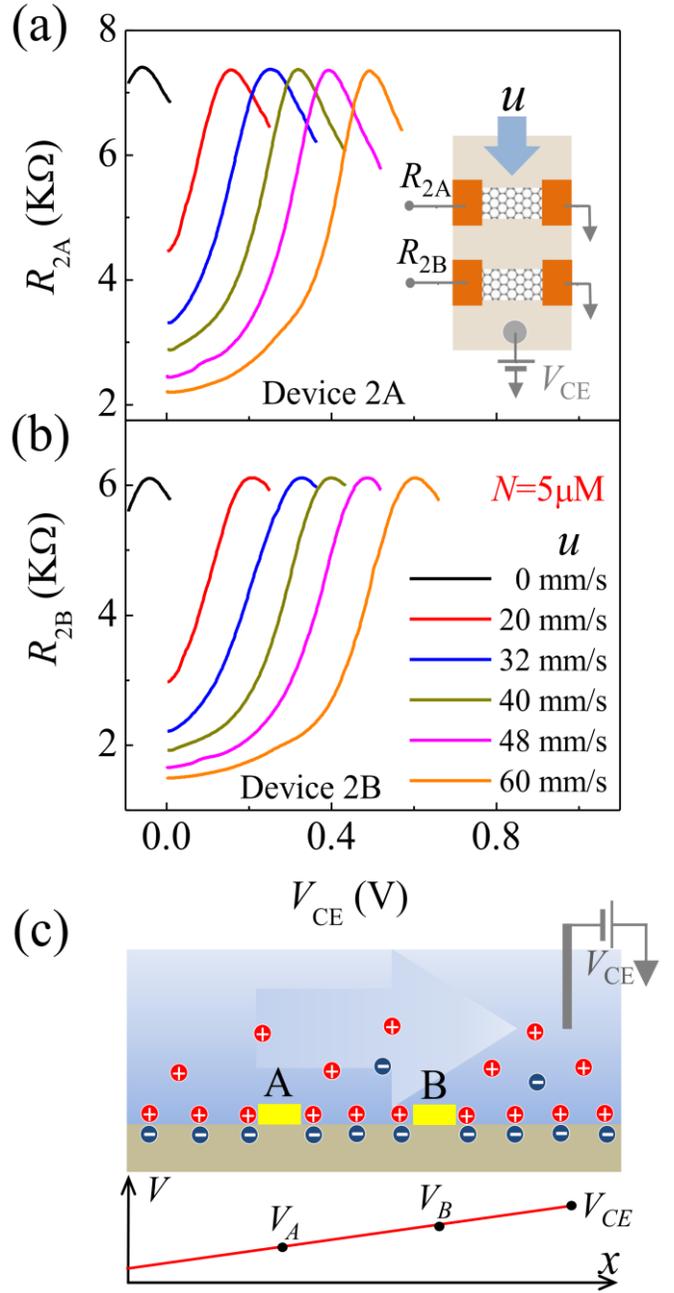

**Figure 2: Flow-dependent transport in GraFETs (Device #2). a)** and **b)** Two-probe resistance vs. counterelectrode voltage $R(V_{CE})$ for different average flow velocities $u$ for two GraFETs (Devices 2A and 2B, shown in the Inset) fabricated on the same chip, located within the same microfluidic channel, and measured simultaneously. The liquid is 5 μM aqueous solution of NaCl. **c)** Top: The schematic diagram showing the ion flow within the diffusive layer of EDL, which results in a streaming potential between points A and B. The counterelectrode is placed downstream from the devices. Bottom: The cartoon illustrating the variation of the local electric potential V along the length of the channel due to streaming potential.

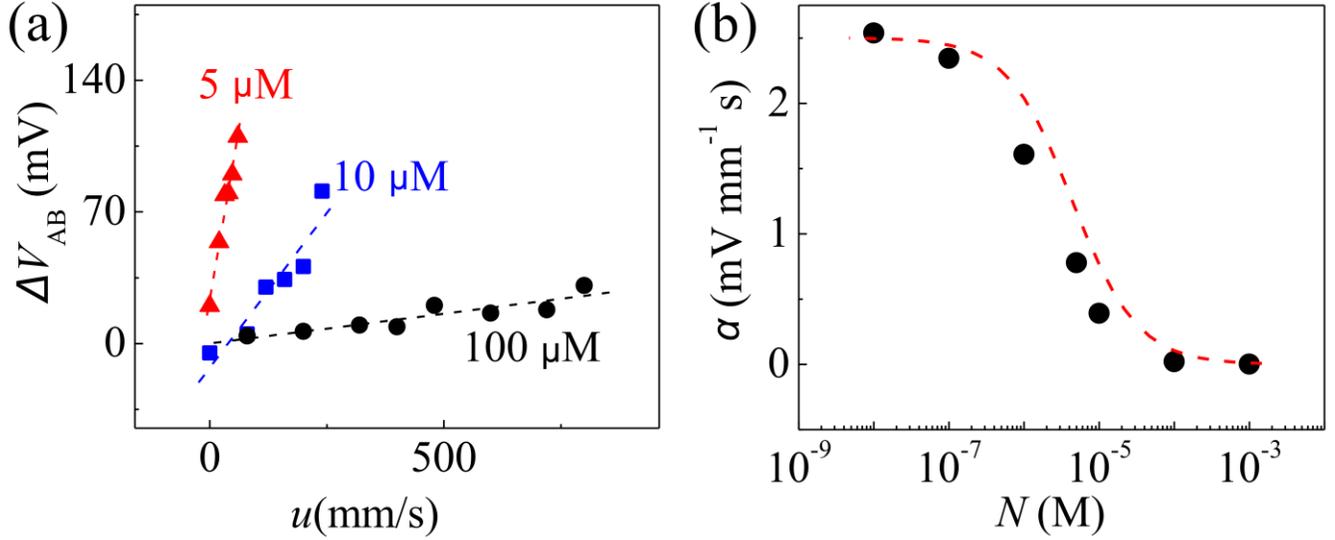

**Figure 3: Analysis of the ionic strength and flow-rate dependent shifts of CNPs in the device #2. a)** The difference between the CNPs of the devices 2A and 2B, $\Delta V_{AB}$, as a function of the average fluid velocity $u$ for three representative ionic strengths $N$ (black circles are $N$=100 µM, blue squares - 10 µM, and red triangles - 5 µM). **b)** The coefficient $\alpha=d\Delta V_{AB}/du$ as a function of the ionic strength. The dashed line is the best fit to the Eq. 1, with parameters $\zeta\sim$-30mV and $\lambda\sim 4\times 10^{-6}$ M.

If streaming potential is indeed contributing to the observed shifts in $R(V_{CE})$ curves, we expect local liquid potential to vary along the channel length. To measure such variation, we fabricated a different specimen (Device #2), where two independently contacted 2-probe GraFETs separated by the distance $L$=800 µm were placed inside the same channel (Devices 2A and 2B in Fig. 2a, Inset and in Fig. 2c). By analyzing the differences in $R(V_{CE})$ of the devices 2A and 2B, we expect to extract precise variation in the local liquid potentials in the immediate proximity to these GraFETs.

Similar to the previously studied Device#1, $R(V_{CE})$ curves for both devices 2A and 2B exhibit reproducible changes with liquid flow (Figs. 2a,b). Importantly, the charge neutrality point of the Device 2B, located further downstream, shifts at a larger rate as compared to the device 2A. To quantify this effect, we plot the relative positions of the CNPs of the devices 2A and 2B, $\Delta V_{AB}=V^{2A}_{CNP}- V^{2B}_{CNP}$, as a function of the flow velocity $u$ and find that this dependence is linear (Fig. 3a). Moreover, the proportionality coefficient $\alpha=d\Delta V_{AB}/du$ is ionic strength ($N$) dependent (Fig. 3a) with maximum being reached at low concentrations (Fig. 3b).

We now quantitatively examine the dependence of $\Delta V_{AB}$ on $u$ and $N$ using a simple electrokinetic theory of streaming potentials based on the Smoluchowski equation.[23, 25] For rectangular channel geometry, the variation of the electrochemical potential over the distance $L$ can be approximated by,[1, 9, 23]

$$\Delta V = \frac{\varepsilon\varepsilon_0 \zeta A R}{\eta e(N+\lambda)\mu_i}u \qquad (1)$$

Here $R\sim 1.3\times 10^{12}$ Pa·s·m$^{-3}$ is the hydraulic flow resistance over a distance $L$ estimated from the Poiseulle's law,[26] $\eta\sim 0.89\times 10^{-3}$ Pa·s is the viscosity of water,[27] $\zeta$ is the electrostatic zeta-potential at the boundary between the compact and the diffusive layer, $e$ is the elementary charge, $\varepsilon$ is the static dielectric constant of water, $\lambda$ is the residual ionic concentration in the liquid due to impurity ions,[23] and $\mu_i\sim 10^{-7}$ m$^2$V$^{-1}$s$^{-1}$ is the effective ionic mobility estimated by measuring the electrical conductivity of the liquid. Treating $\zeta$ and $\lambda$ as fit parameters, we obtain an excellent description of the entirety of our data. The best fit to the data, shown as a dashed line in Fig. 3b, yields values $\zeta\sim$-30mV and $\lambda\sim 4\times 10^{-6}$ M. While the value of the residual ionic concentration for our rectangular channel is reasonable,[1] the obtained value of the $\zeta$-potential is lower than typically reported values for the weighted averages of the zeta potentials of PDMS and $SiO_2 \sim -75mV$.[28] Several reasons can be responsible for this discrepancy. First, while $\zeta$-potential does depend on the ionic strength,[28] it is assumed concentration-independent in our simple model. Second, while most measurements report $\zeta$ for oxygen-plasma-cleaned pristine $SiO_2$ surface, in our experiments oxygen-plasma cleaning was not employed (exposure to ionized oxygen damages graphene devices[17]). We speculate that residues of the electron-beam resist, poly(methyl methacrylate) (PMMA), used in device fabrication still remain on the surface of the channel and can contribute to the observed decreased values of the $\zeta$-potential. Finally, we note that while the Equation 1 was derived for microfluidic channels with aspect ratios <<1, this ratio was close to 1 in our geometry. As a result, the modeling using Eq. 1 can underestimate the zeta-potential.

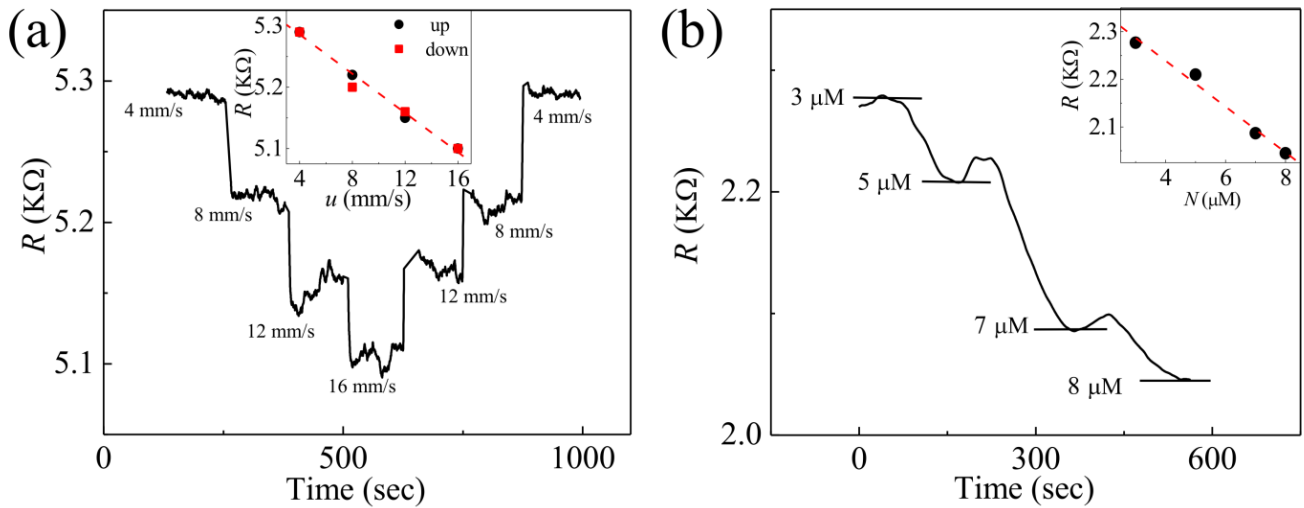

**Figure 4: Figure 4: Performance of the GraFET as mass flow and ionic strength sensor. a)** Mass flow sensing. Resistance vs. time for a GraFET biased at $V_{CE} \sim -0.25V$ as the flow rate is stepped up sequentially through the flow rates of 4, 8, 12, and 16 mm/sec and then backed down sequentially through the same values. **Inset:** Measured resistance as a function of the flow velocity. **b)** Ionic strength sensing. Resistance vs. time for a GraFET biased at $V_{CE} \sim 0V$ as the ionic strength is stepped sequentially through the values from 3μM to 8μM. **Inset:** Measured resistance as a function of ionic strength.

Finally, we use the demonstrated dependence of electrical transport in graphene on the liquid flow parameters to enable precise sensing of mass flow and ionic strength of water. To create both types of sensors, we employed a two-probe CVD-grown GraFET biased at a constant counterelectrode voltage $V_{CE}$, and observed the variation of the resistance with changing mass flow and ionic concentration of a moving fluid (Figs. 4a,b). The resistance of the device was found to be linearly dependent on both the mass flow (Fig. 4a, Inset) and the ionic strength (Fig. 4b, Inset). Consistent with the prediction of the Eq. 1, the maximum mass flow sensitivity was achieved for low ionic strengths.

The sensitivity of the proposed device is determined by the noise present in the system and is ultimately limited by the thermodynamic noise. In our devices, however, the noise in the resistance measurements dominates thermodynamic noise. We estimate the standard deviation of resistance noise in our measurements to be $\sigma^R_{SD} \sim 2\Omega$. This translates into the limit of detection (LOD) of flow rate in our devices (the value that can be detected at $\sim 70\%$ confidence level), $LOD \sim \sigma^R_{SD} \times du/dR \sim 0.1$ mm/s ($Q \sim 25nL/min$). In our experiment, we detected flow rates that are close to this value, $Q \sim 70nL/min$. To put it in perspective, GraFET-based fluidic sensor is $\sim 300$ times more sensitive compared to the previously reported carbon nanotubes[1] fluidic sensors and $\sim 4$ times more sensitive compared to Si nanowires[3] devices.[29] Using similar techniques, we estimated the ionic strength sensitivity of GraFETs to be $\sim 40nM$ (Fig. 4b).

In conclusion, GraFETs, due to the unique ability to sense minute changes in their immediate microenvironment, can be used as powerful probes of electrochemical phenomena in moving liquids. The sensitivity of electrical transport in GraFETs to the velocity and ionic strength of fluids is well explained by the variation of the local electrochemical potential of the liquid next to graphene. We expect that the demonstrated graphene-based label-free sensors for flow and ionic strength may find uses in application ranging from analytical chemistry to bio-molecular detection. Multiple advantages of graphene - the ability to fabricate large area single-layer films of graphene on industrial scale,[30] the possibility to integrate graphene with CMOS processes,[31] high mechanical strength[32] and transparency of graphene[33] – should further contribute to the rapid development of such sensors.

*Acknowledgements:* KIB acknowledges the support through NSF CAREER grant DMR-1056859 and DAM acknowledges support through NIH R21CA126728-01A1. We thank Ronald S. Reiserer, Hiram Conley, J. H. Dickerson, and VIIBRE for technical assistance and useful discussions.